\documentclass[manuscript,natbib=false,nonacm]{acmart}

\usepackage[style=numeric,backend=bibtex,sorting=none]{biblatex}
\usepackage{makecell}
\usepackage{graphicx}
\usepackage{subcaption}
\usepackage{longtable}
\usepackage{array}
\usepackage[inline]{enumitem}  
\usepackage{amsmath}
\usepackage{pdflscape}
\usepackage{svg}
\AtBeginDocument{%
  }

\renewcommand\footnotetextcopyrightpermission[1]{} 
\setcopyright{none}

\begin{document}

\title{Beyond Abstract Compliance:  Operationalising trust in AI as a moral relationship}

\author{Lameck Mbangula Amugongo}
\email{amugongol@gmail.com}
\orcid{0001-6468-2643}
\affiliation{%
  \institution{ Boehringer Ingelheim Pharma GmbH \& Co. KG}
  \streetaddress{Birkendorfer Str. 65}
  \city{Biberach an der Riss}
  \country{Germany}
  \postcode{88400}
}

\author{Tutaleni Asino}
\affiliation{%
  \institution{Carnegie Mellon University}
  \streetaddress{}
  \city{ Pittsburgh, PA}
  \country{USA}}
  \postcode{5000}
\email{tutaleni.asino@gmail.com}

\author{Nicola J. Bidwell}
\affiliation{%
  \institution{Rhodes University}
  \streetaddress{Makhanda}
  \country{South Africa}}
  \affiliation{%
  \institution{International University of Management}
  \country{Namibia}}
    \affiliation{%
  \institution{Charles Darwin University }
  \country{Australia}}
\email{nic.bidwell@gmail.com}

\renewcommand{\shortauthors}{Amugongo, Asino and Bidwell}

\begin{abstract}
Dominant approaches, e.g. the EU’s “Trustworthy AI” framework, treat trust as a property that can be designed for, evaluated, and governed according to normative and technical criteria. They do not address how trust is subjectively cultivated and experienced, culturally embedded, and inherently relational. This paper proposes some expanded principles for trust in AI that can be incorporated into common development methods and frame trust as a dynamic, temporal relationship, which involves transparency and mutual respect. We draw on relational ethics and, in particular, African communitarian philosophies, to foreground the nuances of inclusive, participatory processes and long-term relationships with communities. Involving communities throughout the AI lifecycle can foster meaningful relationships with AI design and development teams that incrementally build trust and promote more equitable and context-sensitive AI systems. We illustrate how trust-enabling principles based on African relational ethics can be operationalised, using two use-cases for AI: healthcare and education.  
\end{abstract}



\maketitle

\newpage

\section{Introduction}
An increasing range of tools, such as checklists, certification schemes, fairness metrics, explainability toolkits and regulatory frameworks have been designed to support developing trustworthy AI \cite{prem2023ethical, deck2024critical, mora2021traceability}. Tool design often aims to implement ethical principles that have broad cross-national consensus, such as UNESCO’s 2021 Recommendation on Ethics of AI, which was adopted by 193 states \cite{van2023ethics}, and instantiate procedural safeguards with value in different contexts, such as audit trails and redress mechanisms \cite{kattnig2024assessing}. Nonetheless, all tools and technical artefacts are shaped by the epistemic commitments and normative power \cite{bentotahewa2022normative} of the regions that dominate AI because they are entangled in their policies, governance structures, institutions, standards, adoption patterns, measurement strategies, datasets, and experimental norms. Thus, tools to support trust by design inherently encode certain epistemologies about trust. 
In this paper, we propose that sensitising tools for trustworthy AI to different epistemologies about trust requires frameworks that emphasise nuance. Replacing entire toolchains is infeasible, and adding yet another distinct framework for trust to the burgeoning range has a limited chance of impact. However, some tools can be redesigned and locally appropriated, rather than assumed to be permanently Western. For instance, cultural prompting and related alignment techniques can reduce Western bias in large-model outputs \cite{tao2024cultural}. Further, the increasing appetite for reflexivity and participatory, context-sensitive practices, amongst the trustworthy-AI community, critical literature and standards bodies, can be guided to sensitise the use of existing tools with more nuanced practices.

In this paper, we focus on Ubuntu, an African philosophy that has received some attention in the field of HCI over the past 20 years, and in Ethical-AI more recently (e.g. \cite{Birhane2021, ewuoso2021african, odero2023ubuntu, gwagwa2022role, okyere2023place, mahamadou2024ubuntuhealth, Amugongo2023, yilma2025ethics, bidwell2021decolonising,bidwell2010ubuntu}). Inspired by previous works, we explore how Ubuntu philosophy contributes nuance to the treatment of trust in AI. We emphasise the importance of integrating what has worked well at the international research ethics level with indigenous African perspectives to build a more robust and holistic understanding of trust. The trust-by-design principles, we propose, enhance relevant trust frameworks by emphasising relational trust, as articulated through Ubuntu philosophy, which addresses some of the limitations associated with individualistic approaches to trust in the international AI discourse. Lastly, shifting from theory to practice, we provide practical steps that illustrate how our trust-by-design principles can be operationalised within the agile development lifecycle.

From here onward, our paper is structured as follows. First, in section \ref{related work}, we summarise how tools for trust by design encode certain Western liberal perspectives on trust, and recent calls in the literature for accounting for meanings about trust that emphasise communal relations. Several authors draw attention to Ubuntu philosophy and explain how approaches in AI design undermine trust relations from an Ubuntu perspective. However, these treatments of Ubuntu trust in AI literature tend to be somewhat superficial; thus, we conclude our literature review by synthesising African work beyond the confines of the AI discourse that honours the nuance. Then, we describe the principles needed to ensure existing tools are sensitised by Ubuntu-trust. Next, we describe two cases, in healthcare and education, that depict how principles can be put into practice.

\section{Related work}\label{related work}
\subsection{Dominant approaches to trust in AI: Trustworthiness as compliance}
Contemporary discourse on trust in AI is largely shaped by regulatory, policy, and technical frameworks that conceptualise trust as a property of systems that can be engineered, measured, and certified. Prominent in these is the European Commission’s Trustworthy AI framework, which defines trustworthiness in terms of lawfulness, ethical alignment, and technical robustness. Similar approaches appear in guidelines from the OECD\cite{oecd2019}, UNESCO\cite{unesco2022}, and national AI strategies \cite{au2024strategy,australiangov2024}, all of which emphasise principles such as transparency, fairness, accountability, privacy, and safety. The growing ecosystem of technical, procedural, and governance tools that support developing AI systems in complying with these principles essentially positions trust as an outcome of regulatory conformity rather than lived experience \cite{euai2024}.

A compliance-oriented approach to trust is supported by auditing tools, metrics, techniques, and documentation practices. Toolkits such as IBM’s AI Fairness 360 and Google’s What-If Tool enable developers to quantify bias and performance disparities across demographic groups, translating ethical principles into computable indicators \cite{bellamy2019ai,wexler2019if}. Model cards and datasheets for datasets aim to prompt and standardise disclosing training data, intended use, and limitations, reinforcing traceability and accountability as documentation practices \cite{mitchell2019model,gebru2021datasheets}, while explainability techniques and model interpretability methods (e.g. SHAP, LIME) are often used to satisfy transparency requirements \cite{molnar2024ai}. AI governance processes, such as ethics review boards, Algorithmic Impact Assessments (AIAs) and risk evaluation checklists, also increasingly institutionalise a compliance-oriented model of trust \cite{reisman2018algorithmic}. 
Principle-based frameworks risk abstraction and superficial adoption. Consider how explainability techniques may satisfy regulatory requirements yet are unintelligible or irrelevant to affected users, which limits their contribution to genuine trust \cite{edwards2017slave,ehsan2021expanding}. Consider also how fairness metrics often involve normative trade-offs that cannot be resolved through technical optimisation alone \cite{selbst2019fairness}. Indeed, organisations can claim ethical legitimacy using principle-based frameworks without meaningfully altering power relations or practices, or “ethics washing” \cite{van2022ai,floridi2019translating,metcalf2019owning}. A compliance-oriented approach to trust not only privileges the perspectives of institutions and developers over users and communities but, in fact, undermines how trust is formed, sustained, or withdrawn. Principles themselves cannot depict how trust is distributed and negotiated in socio-technical practice, and trustworthiness is not a property that systems can have if they align with legal, ethical, and technical standards. Rather, it is a relation that people do or experience, situated in interactions with the real world.

\subsection{Socio-technical and contextual accounts of trust in AI}

In response to approaches that seek to conceptualise trustworthiness as a property that systems can have, scholars increasingly emphasise that trust cannot be abstracted from social relations, institutional histories, and everyday use. Trust in AI systems is conditioned by contextual cues, including interface design, organisational framing, and prior experiences with institutions. For example, transparency features affect trust differently depending on users’ goals and expectations \cite{kizilcec2016much}, and explanations are meaningful only if they align with users’ social roles \cite{ehsan2021expanding}. Drawing on socio-technical traditions in the fields of human-computer interaction (HCI) and Science and Technology Studies (STS) \cite{suchman2007human}, such accounts conceptualise trust as situated, negotiated, and contingent. Socio-technical accounts position trust as emergent in relational work. People do not trust an AI system alone, but rather trust socio-technical assemblages of institutions, professionals, and accountability mechanisms in which users, designers, institutions, and infrastructures must participate in ongoing care. For instance, workers often act as mediators who interpret, contest or compensate for algorithmic outputs \cite{Dahlin2025, disalvo2024workers}. Socio-technical accounts are supported by insights from Critical AI ethics, which show that people may rationally distrust AI systems due to experiences of surveillance, discrimination, and exclusion, regardless of formal compliance with ethical principles \cite{benjamin2023race}. From this perspective, mistrust is not a failure to be corrected but a legitimate response to structural injustice. 

\subsection{Relational ethics and trust as a moral relationship}

Relational ethics offers a profound reorientation of trust in AI and a challenge to instrumentalist, compliance-oriented approaches. Relational ethics argues that trust cannot be reduced to rational choice, calculated risk or probabilistic expectation, but is a form of accepted vulnerability grounded in moral relationships \cite{baier1986trust}. Contemporary ethical theory extends this view by emphasising that trust presupposes normative expectations about the trustee’s motives, character, and responsibilities \cite{hawley2014trust,baghramian2020vulnerability}. To trust is to enter a relationship that carries ethical obligations, including honesty, commitment, responsiveness, and respect. Encouraging the design of trustworthy displaces trust from human actors to technical artefacts, which can not only obscure the moral responsibility of humans and institutions involved, but also create ethical ambiguity \cite{coeckelbergh2020ai}. Relational accounts propose we should explicitly re-frame trust in AI as trust in relationships mediated by AI. User interactions with AI systems instantiate indirect relationships between users and all other stakeholders, in some capacity, including developers, deployers, lawmakers and so on. Thus, if design choices encourage over-reliance or emotional attachment by humans, which lead to deception and manipulation \cite{sharkey2021we,liao2022designing}, designers are implicated in violating norms of respect and honesty. Relational approaches foreground asymmetries of power and vulnerability. Many AI systems are deployed in contexts where users have limited choice, voice, or recourse, and place moral demands on the vulnerable, not commitments undertaken by the powerful \cite{o2020questioning}. 

Feminist ethics of care and relational autonomy have been particularly influential in articulating trust as situated, ongoing, and sustained through relationships rather than guarantees. Thus, overcome modern individualism and replace it with a more relational approach to ethics \cite{Mannering2020}. These traditions emphasise attentiveness, reciprocity, and long-term engagement, offering a sharp contrast to compliance-driven approaches to AI ethics. Trust, from this perspective, cannot be imposed through standards or audits alone; it must be cultivated through practices that acknowledge dependency, power asymmetries, and lived experience. However, relational ethics remains underexplored in mainstream AI ethics literature, which continues to privilege individualistic and procedural conceptions of trust. Where relational perspectives do appear, they are rarely translated into actionable principles for AI development or governance, leaving a gap between ethical theory and engineering practice.

\subsection{African communitarian thought and relational trust}
African relational philosophies offer a particularly rich and underutilised foundation for rethinking trust in AI. Communitarian ethical frameworks, often captured through concepts such as Ubuntu, emphasise personhood as constituted through relationships rather than individual autonomy. Moral value is grounded in mutual care, solidarity, and the flourishing of the community over time. In his papers on the Black box problem and African views of trust \cite{ewuoso2023black} and Epistemic (in)justice, social identity and the Black Box problem in patient care \cite{khan2024epistemic}, Cornelius Ewuoso identifies three clinical gaps created by black-box opaqueness, and explains how they undermine trust relations and patient autonomy when communal, relational or identity-based values matter (e.g., religion, family roles). There is a knowledge/information exchange gap between AI, clinician and patient;  a responsibility/accountability gap and an informed-decision-making gap. Explainability alone is necessary, but not sufficient \cite{khan2024epistemic}. Supporting trust requires not only technical transparency and institutional accountability, but also community/epistemic inclusion. Thus, any framework for trust and supportive tools must not only include technical explainers, docs, institutional audits, liability pathways, but, importantly, processes of participation and documentation of who was consulted and how the values of participants were encoded.

\section{Principles for Trust by Design}
\begin{figure}[htbp]
    \centering
    \includegraphics[width=0.8\linewidth]{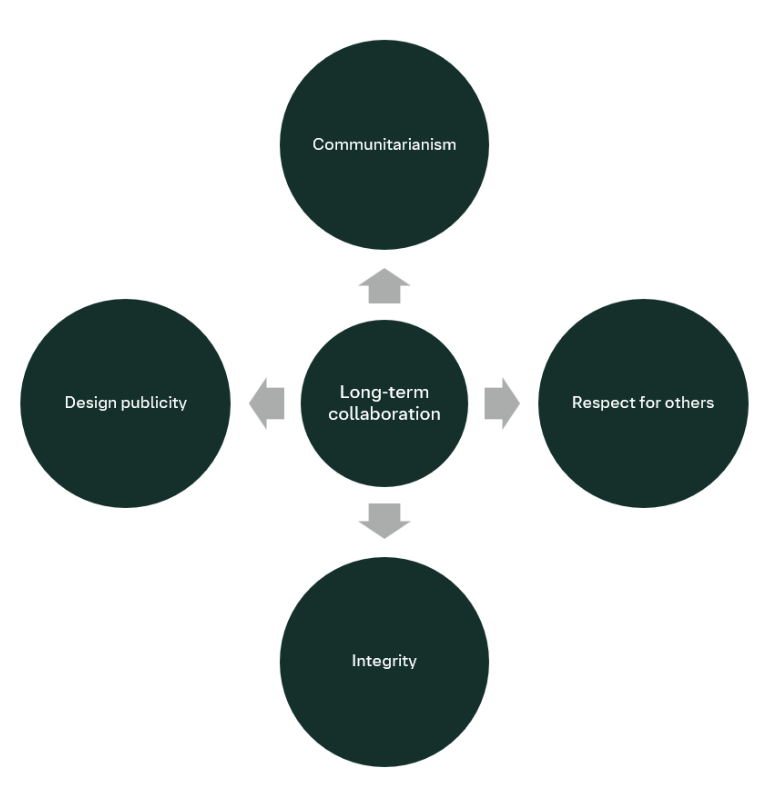}
    \caption{Trust-by design principles centred on long-term collaboration with communities}
    \label{fig:figureFrame}
\end{figure}

In this paper, we propose a set of principles to foster trust in AI by design. Our proposed approach emphasises creating long-term collaboration with communities. We focus on long-term collaboration because it plays a crucial role in fostering long-term relationships and, therefore, trust. Trust and collaboration can be modelled as both independent and antecedent variables. This makes trust an important feature in relationships \cite{AndersonLodishWeitz18987}.  Our Ubuntu-inspired principles focus on 4 principles: communitarianism, Respect for others, Integrity and Design publicity, see Figure \ref{fig:figureFrame}. An overlap may exist between the principles we propose and those found elsewhere. However, we posit the nuances of relational ethics, we emphasise, are essential for fostering trust. We decided on these principles because they focus on prioritising the well-being of the community and ensuring that AI systems do not undermine Community trust and cohesion.

\subsection{Communitarianism}
Existing discourse on AI ethics is largely dominated by values influenced by Western ethical traditions, particularly Kantianism and other liberal philosophical concepts. Normatively, Western norms often prioritise individual rights and autonomy over considerations of the community's good and well-being \cite{Goffi2023}. While such perspectives have shaped much of the global conversation on ethical AI, they risk overlooking the collective dimensions of moral life that are central to many non-Western traditions. In relational societies, communities play a crucial role in shaping the values, morals, and beliefs of individuals. Ethical and moral norms are not formed in isolation but are embedded within social contexts that influence how people understand responsibility, solidarity, and justice. The concept of community has been defined in various ways, with many definitions emphasising geographic space and place. For example, MacQueen et al.\cite{MacQueen2001} identify a core definition of community as “a group of people with diverse characteristics who are linked by social ties, share common perspectives, and engage in joint action in geographical locations or settings.”
Expanding on the definition by MacQueen et al.\cite{MacQueen2001}, we define community as a group of people interlinked through social ties, whether situated in the same locality, connected through religious or ethnic identity, or bound together by shared practices of solidarity and responsiveness. This broader definition acknowledges that communities are not only spatially situated but also socially and culturally constituted. Importantly, we emphasise the nature of African communities—characterised by communal living, collectivism, communal culture, and shared values, which stand in contrast to Western societies that are more strongly characterised by individual values and competition \cite{Mugumbate2023}. This communitarian orientation emphasises mutual care, collective responsibility, and the prioritisation of the common good over individual gain. 
Scholarships on communitarianism were brought to the forefront by thinkers such as Michael Sandel and Michael Walzer. These two scholars represent divergent views of contemporary communitarian thinking. For example, Sandel’s communitarian views emphasise civic republicanism, striving to achieve a normative goal of promoting virtue and participation of citizens in their self-governance \cite{Sandel1982}. On the other hand, Walzer emphasises a more egalitarian approach, whereby normative goals of what is right and just are intrinsically connected to communities’ shared beliefs \cite{Walzer2007}. While influential, these perspectives do not encompass the entirety of communitarian thought, which varies significantly across international contexts and philosophical concepts.
Beyond Western debates, communitarian concepts exist in diverse international contexts. For example, Chinese communitarianism, deeply rooted in Confucianism, prioritises social harmony and collective duty over individual autonomy. Hwang \cite{Hwang2011} highlights the principle of benevolence and unboundedness, which Confucius considered the foundation of all human relationships. This principle embodies kindness, generosity, and the encouragement of selfless caring and thoughtfulness, reflecting a communitarian value that privileges relational responsibility. In Russia, there is a concept of sobornost, which can be translated to mean conciliarity, togetherness or unity, introduced by Alexei Khomiakov and later elaborated by Russian philosophers \cite{Khoruzhyi2014Sobornost}. Sobornost emphasises peaceful coexistence that extends not only between individuals but also between people and their broader social and natural surroundings, underscoring a holistic vision of community. In Latin America, communitarian traditions rooted in indigenous philosophies emphasise reciprocity and collective well-being, shaping grassroots movements and national projects that challenge neoliberal individualism and promote collective empowerment \cite{RossetPhanthuwongpakdee2025}.
In Africa, communities have long practised communitarianism, including traits such as solidarity, sharing, caring, mutual interdependence and complementarity \cite{Chemhuru2018}. Though there may be common traits between communitarianism theories, i.e., all emphasising the relationship between community and individuals. Afro-communitarianism provides a more holistic conception of interconnectedness, uniquely fusing communal solidarity with recognition of individual dignity \cite{Chimakonam2022}, affirming that the dignity of every person should be respected regardless of whether they are deemed rational or not. Moreover, Afro-communitarianism emphasises strong long-term relationships built over time, creating trust in the process. Other researchers have shown that Afro-communitarianism, such as Ubuntu, can serve as an effective framework for delivering trusted, empowering, and strengths-based professional social work services within communities \cite{Mugumbate2023}. Similarly, in education, \cite{Chingombe2019} outline that Ubuntu’s inclusivity and protection for those deemed weak help build confidence and trust among learners. As such, we posit that Afro-communitarianism can reshape how we think about trust in AI. By embedding relational trust and Afro-communitarian values into AI development, this paper contributes a novel perspective to AI ethics and governance. It challenges dominant technical and legal paradigms by advocating a normative, participatory, and culturally grounded approach. This shift strengthens trust while promoting equitable, inclusive, and socially responsive AI. Ultimately, sustaining trust in AI requires more than robust algorithms or ethical checklists; it calls for reorienting power, purpose, and process toward the collective well-being of all.

\subsection{Respect for others}
The concept of “respect” is not new in ethics; the principles of “respect for autonomy” have been dominantly used in bioethics. Bioethics principles are the basis for many AI ethical principles and guidelines, such as the AI4People framework \cite{Floridi2018}. From bioethics, respecting autonomy implies that every individual has the right to make their own decisions, e.g., the treatment they do or do not receive. Researchers have argued that “respect for autonomy” is unsatisfactory as it is grounded in an individualistic worldview, failing to acknowledge the fundamental importance of understanding personhood within the nexus of their communal relationships \cite{Behrens2018}.
Behrens replaced the principle of autonomy with the principle of “respect for persons”. He argues that the principle of “respect for autonomy” does not resonate with basic relational moral convictions, for example, relational beings who are embedded in their communities. This critique is shared by other researchers who have argued that non-Western ethics are more suited for less individualistic societies. As an alternative to “respect for autonomy”, “respect for persons” has been proposed as a more appropriate principle, as it acknowledges both communal relations and individual choices \cite{Behrens2018}. In this study, we support the use of “respect for persons” as a more holistic principle to ground AI systems to account for degrees of moral status.
We acknowledge that the definition of personhood has been contentious in bioethics and ethics discourse. For clarity,  we adopt the definition of relational personhood that differentiates between individuals with agency and those without, asserting that our moral responsibilities vary for each group, as clearly outlined by Metz \cite{MetzThaddeus2007TaAm}. We believe this definition better aligns with relational moral intuitions compared to Western intuitions informed by Kantianism. Thus, respect the inherent dignity of every person, regardless of whether a person is deemed to have a reduced rational capacity. For example, AI systems built on data scraped from the internet without data owners’ consent do not adhere to the principles of “respect for persons” and, as a result, violate the data privacy of individuals. For such scenarios, regulations and data protection laws are essential for protecting individuals' right to informed consent \cite{Amugongo2023}. However, existing regulations, such as the EU Act, have broad definitions and, in their earlier version, categorised complex AI systems such as LLMs as chatbots. Novelli et al.\cite{NOVELLI2024106066} highlighted that the EU AI Act is not adequately prepared to deal with the lawfulness and accuracy of unpredictable content generated by LLMs. In addition, LLMs may also be trained with personal data. Going forward, we need a holistic governance mechanism that will ground complex AI systems to balance individual rights and societal interests, clearly outlining metrics for privacy-preserving methods and stipulating options for individuals to opt out of their data from any model. 
Relational ethics, such as Ubuntu, emphasise genuine solidarity with communities and transparent interactions as a way to maintain communal relations. Thus, trust. Lastly, trust, autonomy and respect are related. Therefore, to foster genuine trust in AI, we need to ensure that AI systems perform equitably and respect existing relationships of trust by exhibiting solidarity and protecting combined communal interests beyond individual informed consent. This can be realised by establishing a mechanism for obtaining collective informed consent and community vetting of AI systems that ensure that community common interests are protected.

\subsection{Integrity}
Integrity is an important dimension of trust, playing an important role in influencing users' trust. Integrity-based principles include honesty, transparency and fairness \cite{Mehrotra2024}. A recent study has demonstrated that people evaluate AI chatbots similarly to how they judge other humans, with their trust judgments based on the same psychological mechanisms \cite{Lalot2024}. They highlighted integrity and competence as the important characteristics of trust, as they help humans determine whether an AI is reliable.
Ethics and integrity are intertwined. In the literature on integrity, there is no homogeneous or clear definition of integrity. Generally, terms such as wholeness, completeness, consistency and coherence, moral reflections and principled values are used to define integrity. One of the dominant views of integrity is based on \cite{Montefiore1999}, defining integrity as "wholeness" or completeness involves the consistency and coherence of principles and values. Other views focus on specific values: impartiality, honesty and accountability \cite{Dobel2016}, the connection between integrity and morals, encompassing what is right and wrong, good or bad \cite{Carter1996}, and legal views \cite{rosenbloom2011legal}. 
The legal view of integrity is attractive as it assumes regulations reflect the society's values and morality. In the context of AI, regulations such as the EU AI Act can play a useful role in ensuring that developers of AI systems act in a responsible way and develop safe, ethical technologies, which may instil confidence in AI. However, laws have limitations; they are not always applicable to behaviours being judged and may sometimes contradict societal dominant values on acceptable behaviours \cite{Huberts2018}. Moreover, it is very difficult to enforce laws, especially in behaviours that happen out of public sight. For example, most used LLMs are developed as closed systems, where only institutions that develop them have insight and knowledge of how they work and whether they conform to the rules.
We need moral values and norms as guiding principles for integrity. We posit that integrity as a moral quality will ground those developing AI systems that act in accordance with relevant moral values. Critiques will highlight that there is no one morality. So, whose values should AI conform to for it to be judged as having higher integrity? The dominant view is that AI should be grounded in fundamental human rights to promote accountability. Despite containing well-established principles, human rights frameworks may not fully address the new ethical challenges posed by AI, for example, as shown in predictive policing \cite{Vidushi2020} and content moderation \cite{martin2020participatoryproblemformulationfairer}. But can provide a set of guiding principles for aligning AI values across diverse cultures and contexts, given their qualified cross-cultural validity \cite{prabhakaran2022humanrightsbasedapproachresponsible}.
In practice, integrity in AI development means doing what is morally right, even when not legally required. It involves disclosing training data, documenting quality checks, and maintaining audit trails to ensure accountability. By embedding integrity into the design and governance of AI systems, developers can build systems that are not only technically sound but also ethically and socially credible.
Trust is a by-product of integrity, and integrity is built through honesty, consistency, and alignment between one’s words and actions, which in turn fosters credibility and reliability. Hence, trust in AI is only possible if those who design, deploy, and govern these systems act with integrity in both intention and practice. But trust is not merely an individual attribute; it is relational. It emerges through ongoing interactions, mutual understanding, and the willingness to be vulnerable in the presence of another’s power. In human terms, trust grows through reciprocity and care; in AI, it must be cultivated through transparency, accountability, and responsiveness to those affected by its decisions. Thus, trust is sustained not by perfection, but by the continual repair and reaffirmation of relationships when expectations falter.

\subsection{Design publicity}
The opaque nature of many ML models, particularly deep learning-based algorithms, has raised significant concerns about explainability and interpretability. These concerns are especially pressing in high-stakes domains such as healthcare, criminal justice, and finance, where algorithmic decisions can have profound consequences. Traditionally, explainable AI (XAI) has relied on post-hoc interpretability methods, such as SHAP, to provide insights into feature contributions \cite{Lipton2018}. However, these methods have notable limitations: they may oversimplify complex models, fail to guarantee accurate explanations, and risk misleading users about the true basis of algorithmic decisions \cite{Vale2022, Loi2021}.
To address these shortcomings, design publicity has emerged as a more robust approach to algorithmic transparency. As proposed by Loi et al.\cite{Loi2021}, design publicity entails the proactive communication of essential elements that allow individuals to assess whether an algorithmic decision is justified. This includes not only technical details but also normative justifications, why a model was designed in a particular way, what values it encodes, and how it aligns with legal and ethical standards. Unlike post-hoc explanations, design publicity offers “explanation by design”, embedding transparency into the architecture and governance of AI systems from the outset. We support the adoption of design publicity and extend it by integrating relational ethics, particularly those grounded in African communitarian thought. Relational ethics emphasises inclusivity, mutual accountability, and the moral significance of social bonds \cite{Lansing2023}. In relational societies, trust is not merely epistemic but deeply moral, built on shared experiences, solidarity, and demonstrated goodwill \cite{ewuoso2023black}.
Transparency is essential for maintaining fiduciary relationships, especially in contexts like healthcare, where Clinicians are moral obliged to act in their patients' best interests. If they rely on opaque AI systems, their ability to uphold this fiduciary duty is compromised. As Ewuoso\cite{ewuoso2023black} argues, explainability enhances clinicians’ capacity to act intentionally and transparently, provided they retain control over the AI and consider patients’ values. Relational theories such as Ubuntu further illuminate the connection between trust and trustworthiness. In contrast to Western models that emphasise abstract indicators like competence and integrity \cite{faulkner2017}, relational trust is grounded in personal knowledge, communal reputation, and lived experience. Trust is not just about realising expectations; it is about striving to do so with goodwill and relational accountability.
A major challenge for designing publicity is the proprietary nature of many AI systems. Companies often withhold model details, which they justify in relation to intellectual property law yet undermines public trust and democratic oversight. Design publicity offers a middle ground: rather than disclosing sensitive code, developers can share feature attributions, training data summaries, and value-sensitive design rationales \cite{Gebru2021}. Moreover, current practices such as external red-teaming, used by companies like OpenAI, often exclude the very communities most affected by AI risks \cite{openai2024redteaming,beutel2024diverseeffectiveredteaming}. We argue that community involvement must be institutionalised, not optional. Regulations should mandate collaboration between developers, designers, and impacted communities to ensure that AI systems reflect communal interests rather than commercial imperatives.

Design publicity, when combined with relational ethics, offers a powerful framework for building genuine, trust-based relationships between AI systems and the communities they serve. Transparent AI systems enable users to understand and justify decisions, fostering validation, accountability, and moral legitimacy. By embedding transparency into the design process and grounding it in relational values, we can move toward AI systems that are not only explainable but also trustworthy, inclusive, and aligned with the common good. While common methods like Model Cards \cite{mitchell2019model}, datasheets \cite{gebru2021datasheets}, and audits enable transparency, truly building trust is active community involvement, embracing a feedback loop, shifting accountability by empowering communities to question, review Model Cards or AI audits. Trust in AI is not merely a technical problem; it is a socio-technical challenge that requires collaborative, transparent, and ethically grounded solutions. Design publicity, when combined with relational ethics, offers a powerful framework for building genuine, trust-based relationships between AI systems and the communities they serve. Transparent AI systems enable users to understand and justify decisions, fostering validation, accountability, and moral legitimacy. By embedding transparency into the design process and grounding it in relational values, we can move toward AI systems that are not only explainable but also trustworthy, inclusive, and aligned with the common good. While common methods like Model Cards \cite{mitchell2019model}, datasheets \cite{gebru2021datasheets}, and audits enable transparency, truly building trust is active community involvement, embracing a feedback loop, shifting accountability by empowering communities to question, review Model Cards or AI audits.

\section{Operationalising trust-by-design principles}
\begin{figure}[htbp]
  \centering
  \includegraphics[width=1\columnwidth]{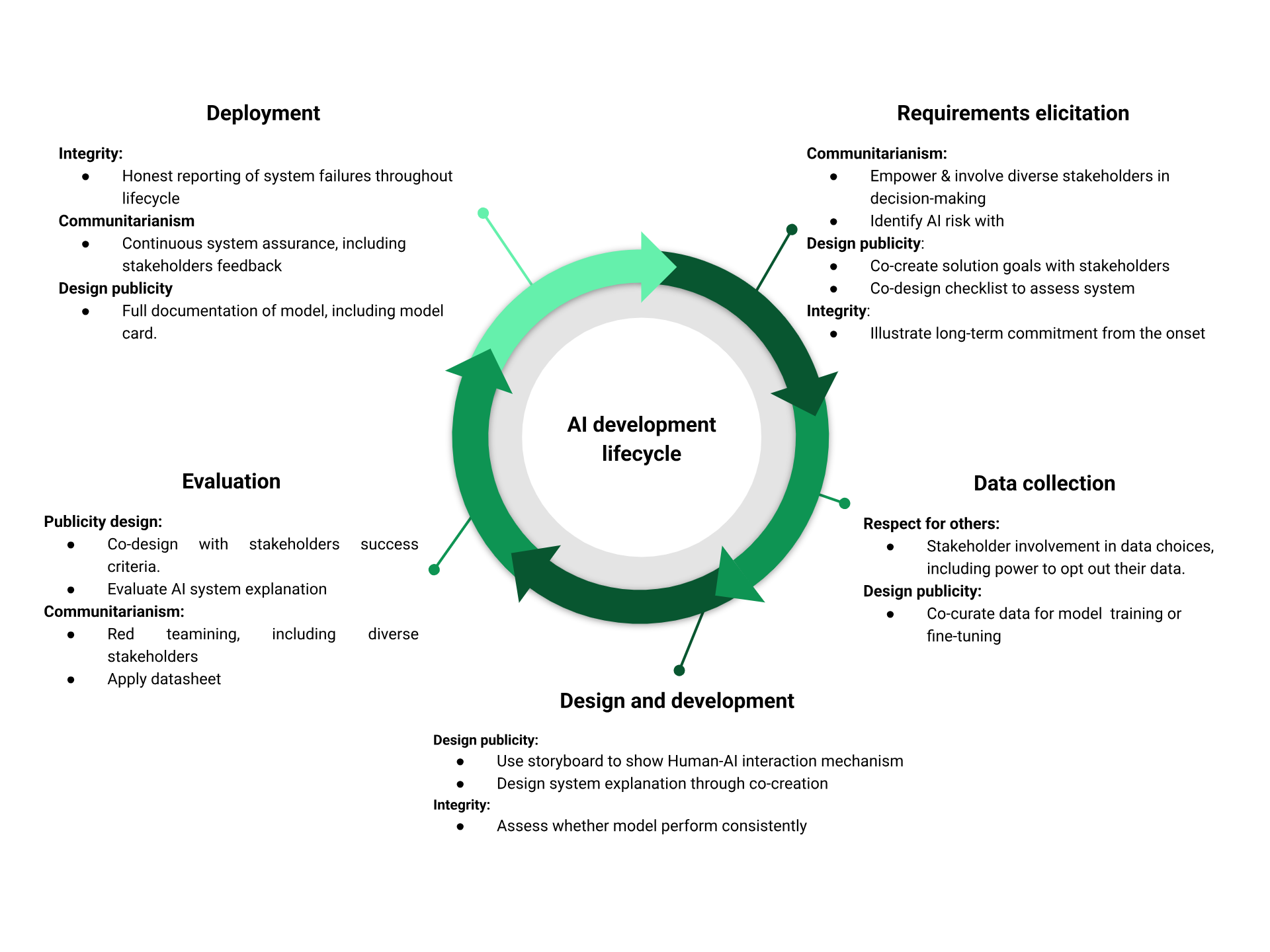}
  \caption{Integrating the trust-by-design principles into the agile AI development lifecycle}
  \label{fig:figureOperationalise}
\end{figure}

Translating ethical principles into practice is not new; previous studies have proposed techniques to operationalise ethical principles. For example, checklists were co-designed with practitioners to address AI fairness concerns \cite{Madaio2020}, operationalising principles across the AI lifecycle and contextualising a framework to empower developers to operationalise ethical principles for M-Health applications \cite{Amugongo2023SDLC}. Inspired by the aforementioned works, we illustrate actionable steps for developers to operationalise our trust-by-design throughout the AI development lifecycle. We use agile software development in our illustration, because it is commonly known amongst AI developers and will make it easier for developers to operationalise than another standalone trustworthy AI framework. Figure \ref{fig:figureOperationalise} illustrates the step-by-step guide of how our trust-by-design framework can be put into practice across the development lifecycle. Informed by relational ethics, our framework emphasises that trust in AI systems must be cultivated as a shared, ongoing process that is embedded throughout the entire AI lifecycle. Trust, as relational ethics suggests, is not a static attribute but a dynamic, bi-directional relationship that requires continuous effort and mutual accountability between the trustor (e.g., the user) and the trustee (e.g., the AI system and its developers) \cite{ewuoso2023black}. This aligns with the principle that trust cannot be retrofitted into an AI system after deployment; instead, it must be intentionally integrated into every phase of the AI lifecycle.

\textbf{Requirements elicitation}: Requirements elicitation is a crucial phase in the development life cycle, where the goal is to identify and establish the functional and non-functional requirements, including ethical requirements. In addition to functional and non-functional requirements, it is essential for the system goal to be clearly defined and stakeholders. As such, during this phase, developers should engage communities as stakeholders in identifying the need goal of the AI system. Involving stakeholders early on will facilitate the trust-building process, which is an antecedent of strong relationships. The communitarianism principle ensures that all stakeholders are involved in key decision-making about the system, addressing the power balance. Trust is a by-product of integrity, built through honesty, consistency, and alignment between one’s words and actions. Our trust-by-design principle emphasises that both developers and communities should exhibit goodwill and interact transparently and honestly from the onset to build trust.

\textbf{Data collection}: Data is the lifeline of any AI system. Before data collection, we recommend a clear data governance principle that does not amplify social inequality but ensures socially beneficial data production and use \cite{Viljoen2021}. Shifting from an individualistic model of data subject rights to a more democratic approach to data governance, where communities co-decide on the data to be used, ensuring that the dataset is diverse and representative of real-world contexts. The principle of respect for others grounds data collection to prioritise dignity, autonomy and equitable fairness.

\textbf{Design and development}: Studies from HCI have shown the importance of and how co-designing systems with users helps build trust \cite{Wang2019XAI, Zicari2021, Auala2025}. In the context of AI, Wang et al.\cite{Wang2019XAI} co-designed a human-centred, decision-driven approach through XAI with clinicians, improving explainability and trust. In another study, Zicari et al.\cite{Zicari2021} showed that co-creating AI systems improves the trustworthiness of AI. During the design phase, we recommend co-design and co-creation sessions with stakeholders to formulate how the system will work through storyboarding, system interaction mechanisms and co-design evaluation metrics with the user. Co-creation of the system design ensures that the system respects community and cultural sovereignty, at the same fostering "a sense of ownership and to ensure that the representations align with local values and lived experiences" \cite{Auala2025}.

\textbf{Evaluation}: Assessing how well an AI system performs against the set goal is an important measure to assess system performance. Quantitative metrics using benchmarks are commonly used to assess system performance. However, quantitative metrics can reduce complex AI system operations to numbers. AI systems are socio-technical systems; as such, we need diverse perspectives to holistically evaluate system performance beyond benchmarks. During the evaluation phase, the design publicity principle enforces that evaluation success criteria should not be imposed solely by developers but co-defined with stakeholders. To operationalise trust during evaluation, it is crucial to move beyond legal and technical safeguards and approach trust as a dynamic, relational process. This involves co-designing benchmarks that are beyond Q\&A benchmarks. Evaluation metrics should reflect transparency, mutual respect, and long-term engagement, validated through participatory methods that include community feedback.

\textbf{Deployment}: Building trust in AI requires addressing risks across the entire AI lifecycle. While regulation plays an important role in preventing harm and ensuring accountability, we recommend that AI developers adopt a proactive approach that treats AI as a shared responsibility throughout its use. During deployment, periodic audits should be conducted to assess whether model recommendations continue to align with established guidelines. Version control is particularly challenging for LLM-based systems, where updates are less transparent than in traditional software. For example, if an update introduces inaccurate outputs, mechanisms should enable rollback to the last stable model version to preserve reliability. Transparent reporting mechanisms should also be established for system failures or biases, such as disproportionate antibiotic recommendations for certain patient groups without clear justification. Finally, accountability must be enforced across the AI lifecycle through clearly defined responsibilities and escalation protocols for failures. This can be supported by a multidisciplinary oversight committee representing all stakeholders, tasked with periodically reviewing system performance to ensure accountability is shared and decisions reflect community values.

The aforementioned steps provide general consideration to operationalise our trust-by-design principle. Gillespie et al.\cite{Gillespie2023} found that people’s trust in AI may vary depending on the application of AI. Similarly,  Amugongo, Bidwell, and Mwatukange \cite{amugongo2025trust} found that in a survey among Africans with educational interest in AI, trust in AI systems varies significantly by domain. In both studies, participants have highlighted that they are more trusting of AI in healthcare compared to AI that uses personal data. 
Acknowledging that people’s trust in AI is fundamentally dependent on the application domain, with different use cases, such as high-risk healthcare applications versus lower-risk educational tools, generating varying levels of user expectation \cite{Gillespie2023,amugongo2025trust}. Our proposed trust-by-design approach emphasises the universal applicability of relational trust theories, which ground the trust-building process in consistent norms irrespective of the specific use case. Specifically, we draw upon relational ethics, particularly as articulated in the philosophy of Ubuntu, which emphasises the necessity of cooperative behaviours and mutual accountability in all interactions. To consistently exhibit the virtue of cooperativeness, a cornerstone of trust, AI systems and those developing them must embody specific attitudes such as transparency, honesty, enable users to make voluntary choices, act as a trustworthy partner, and align with common goals. We will illustrate how these principles apply across various contexts by providing contextualised examples that utilise AI in the medical field and in education. These relational norms are critical to integrating trust into AI systems, ensuring they are not only technically robust but also contextually aligned with the core values and expectations of the communities they serve. We will illustrate how these principles apply across contexts by showing contextualised examples using AI in the medical field and AI in education.

\subsection{Scenario 1: AI in healthcare - AI-enabled decision support for Antibiotic prescription}
Healthcare is a high-stakes environment, and trust is crucial for acceptance and adoption of AI systems. This is primarily because clinicians have a moral obligation and fiduciary duty to promote patients' best interests. As stated by Beauchamp and Childress: “The patient-physician relationship is a fiduciary relationship, that is, founded on trust or confidence; and the physician is therefore necessarily a trustee for the patient’s medical welfare” \cite{Beauchamp2013}. As such, it is important for AI systems used in healthcare to aid clinicians in maintaining this fiduciary responsibility.

During the requirements phase, we recommend integrating the principles of communitarianism and Design publicity to foster trust from the onset. Strong continuous relationships and human oversight have been found as factors that affect trust in AI in healthcare \cite{Astobiza2025}. Communitarianism involves engaging all relevant stakeholders: patients, clinicians, pharmacists, and developers as co-creators rather than passive end-users, which helps build relationships and subsequently build trust through shared responsibility \cite{Kaye2024}. For example, hosting multi-stakeholder workshops can surface diverse perspectives on antibiotic prescribing challenges, such as the balance between rapid therapy and antimicrobial resistance concerns. Framing system goals to be aligned with potential patient needs through problem co-creation, as shown in a study by \cite{Stawarz2023}, will ensure relevance and alignment with existing, trusted clinician-patient relationships. Design publicity ensures this process is transparent, allowing stakeholders to jointly define system goals, such as reducing inappropriate prescriptions without compromising outcomes. These co-creation sessions can also identify potential system risks, such as over-reliance or bias, and develop shared ethical checklists for fairness and explainability, making design choices publicly justifiable and fostering initial accountability and trustworthiness \cite{Stawarz2023}.

During the data collection phase, we integrate the principles of Design publicity and respect for others to foster trust and ensure socially beneficial outcomes. Adopting a democratic, communitarian approach shifts the focus from individualistic data rights to collective governance, where communities and clinicians co-decide on the data to be used \cite{Viljoen2021}, ensuring the dataset is diverse and representative of real-world contexts. This ensures the dataset, which may include local resistance patterns and clinical histories, avoids hidden biases and reflects diverse clinical realities. The principle of respect for others can be upheld by partnering with the community to vet proposed personal data elements, like previous prescription records or comorbidities, ensuring relevance and necessity while prioritising dignity and autonomy. For example, consent processes will go beyond a one-time checkbox, allowing patients granular control over data reuse without penalty. This participatory approach, which includes involving stakeholders in training dataset curation and communicating system data limitations, ensures transparency and helps maintain strong relationships between developers and the community, thereby fostering trust in the system.
As a relational concept, transparency is an antecedent for fostering genuine relationships, therefore, trust \cite{ewuoso2023black}. During the design and development phases, integrating the principles of "design publicity" and a communitarian approach is necessary for fostering trust. To address the challenge that typical XAI explanations often rely on developer intuition rather than user needs, our approach emphasises co-designing the explanation with stakeholders, clinicians, patients and communities. This communitarian engagement ensures that the innovation is viewed from the perspective of lived experience, not just technical expertise, creating truly community-centred AI. For example, during co-design workshops, stakeholders can co-design the system using real-world scenarios and provide feedback on whether explanations align with clinical guidelines and individual patient contexts. This iterative refinement ensures explanations are not only informative but also help clinicians maintain their fiduciary responsibility to patients. Documenting these design decisions, such as why a particular explanation format was chosen, upholds the principle of design publicity by creating an auditable trail that supports accountability. This transparency maintains a strong relationship between system developers and the community the AI system intends to serve, ultimately building the necessary trust for effective deployment.
During the evaluation phase, we integrate design by publicity and communitarianism principles to ensure a comprehensive assessment that goes beyond quantitative benchmarks. Design publicity ensures evaluation success criteria are co-defined with stakeholders through participatory workshops. In this workshop, clinicians, pharmacists, and patient representatives can collaboratively define success, e.g., reducing inappropriate antibiotic use without increasing treatment delays, while also tracking metrics for unintended consequences like alert fatigue. This participatory approach moves beyond developer-imposed metrics and ensures evaluations reflect the healthcare community's collective values and priorities.
The red-teaming process aims at discovering AI vulnerabilities and flaws, involving red-teamers designing threat models \cite{openai2024redteaming,beutel2024diverseeffectiveredteaming}. The inclusive and collaborative nature of relational ethics emphasises that stakeholders should be involved in the threat model design process.  For example, one potential threat model for an AI-enabled decision support for antibiotic prescription could describe how the system is vulnerable to recommending unnecessary or incorrect antibiotics when prompted by a user misrepresenting a patient's symptoms, causing patient harm and contributing to antibiotic resistance. Relational ethics can shape the red-teaming process by shifting the focus from technical failures to the impact on the relationships of care between patients, clinicians, and the healthcare system. Rather than just testing if the AI provides the correct recommendation. Red-teaming scenarios should explore power imbalances and vulnerabilities in the clinical workflow. For instance, red teamers might explore if the AI's recommendations could be used coercively or if it consistently under-serves specific patient populations. By incorporating these social and relational dynamics, the red-teaming becomes more inclusive, addressing a wider range of potential harms. This transparency and proactive consideration of human-centred vulnerabilities can ultimately increase trust in the AI among both clinicians and patients, ensuring it serves as a supportive tool rather than a source of potential friction or harm within the patient-clinician relationship.
Accountability mechanisms are easier to set up when stakeholders are transparently aware of factors that influence model decision-making. During the deployment of an AI-enabled decision support for antibiotic prescription, design, publicity and integrity are important principles for building trust with clinicians and patients. Design publicity involves transparently sharing information about the system, such as what data the model was trained on and how it influences antibiotic recommendations. For example, making a dashboard available that shows the system's performance, including AI failure rates, to help doctors understand its reliability and limitations. The principle of integrity means that all stakeholders, including the AI developers and clinicians, act with the intention to do good and will take accountability for any harm caused. If the system suggests an inappropriate antibiotic that leads to a patient's adverse outcome, the team responsible must acknowledge the failure, investigate the cause, and work to repair the damage to maintain the trust relationship with clinicians and patients. Demonstrating a commitment to acting in the best interest of the patient, reinforcing the trust that the AI is a benevolent partner in care. Operationalising our trust-by-design principles in this context requires embedding relational ethics into the AI lifecycle to ensure that the system supports clinicians in making informed, ethical, and culturally sensitive decisions. This is important because clinicians have a fiduciary responsibility towards patients, and this fiduciary responsibility cannot be delegated to AI. As such, AI systems applied in healthcare should aid Clinicians to uphold this fiduciary responsibility. The principle of design publicity emphasises co-creation with communities, allowing stakeholders to meaningfully influence design decisions. For example, co-design explanations with stakeholders will ensure explanations are not only informative but also help clinicians maintain their fiduciary responsibility to patients.

\subsection{Scenario 2: AI in Education}
In contrast to healthcare, applications of AI in education may be perceived as a lower-risk application of AI. However, trust in this scenario implies that AI should respect learners’ dignity, adapt to inclusive pedagogies, and avoid reinforcing educational inequalities, i.e., bias and stereotypes. A study by \cite{Hlatshwayo2020} found that the application of Ubuntu principles in higher education builds trust, confidence, and relational learning for both students and teachers. So, during the requirements elicitation phase, communitarianism principles emphasise that stakeholders should be involved in defining the goal. In the case of an AI system in education, the goal is to create an AI tutor personalised for every learner, not to replace teachers. If the goal is to replace teachers, it fundamentally undermines relational bonds, reducing education to transactional knowledge transfer rather than a communal act of growth. Thus, contradicts exhibiting solidarity and collective well-being, effectively causing harm by isolating learners from the human empathy and trust that teachers embody. Trust cannot be cultivated when relationality is removed, because Ubuntu views education as a shared journey of becoming together, encompassed in phrases such as “it takes a village to raise a child”.
In South Africa and Namibia, there is a lack of social cohesion, coexistence and cooperation amongst learners, which stems from apartheid, which divided people according to racial and cultural lines. If an AI system is to be developed to be used in this context. Relational theories emphasise collective responsibility to ensure that AI used in education fosters collaborative learning, cultural sensitivity, and social justice constitute foundational imperatives in education \cite{MulaudziGundo2024}. Any deviation from these principles represents a failure of the moral and professional duty to construct equitable and sustainable educational systems that foster prosperity for all. During data collection, choices such as pedagogical and data choices should be inclusive, upholding the principle of respect for others by acquiring consent for data and ensuring data sheets are used to document the dataset. Moreover, ensure that the data is diverse with content in different languages to allow learners to be able to engage in their indigenous languages, ensuring that no learner is left behind.
During the design phase, applying Ubuntu principles will challenge the dominance of Western epistemologies by placing Indigenous knowledge systems at the heart of curriculum design \cite{Ngubane2021} and ensuring inclusive pedagogy is applied. Relational principles of communitarianism, integrity and collective well-being will ensure equitable access to quality education, creating personalised learning experiences that actively dismantle long-standing systemic barriers for marginalised communities \cite{Suliman2024}. Teachers, students, and parents should be involved in co-designing learning pathways. For example, AI tutors should reflect local curricula and linguistic diversity, ensuring relevance and inclusivity. The principle of respect for others will ensure that the system prioritises inclusivity by avoiding shaming learners for mistakes and instead promoting a growth mindset, and incorporating feedback mechanisms that are culturally sensitive and affirming, fostering confidence and motivation among students.
Technology reproduces stereotypes and reshapes cultural meaning \cite{Mokoena2023UbuntuEthicsTechnology}. Additionally,  the perpetuation of Western cultures can reinforce existing power structures, which makes AI systems biased. During the evaluation phase, the model should be assessed using different approaches, including normal white-box and black-box testing and red teaming. Red teaming exercises can involve stakeholders simulating misuse scenarios, such as over-reliance on AI grading or inappropriate feedback. This should not be a once-off process but should be periodically assessed to identify gaps and failures of the system throughout the development lifecycle. As such as during deployment, the principle of integrity emphasises that the system operations are audited and the community is transparently informed about system failure.
In conclusion, relational theories such as Ubuntu will not only address the systematic barriers and epistemic biases entrenched in education that have historically excluded marginalised communities, but also foster trust in AI systems \cite{Hlatshwayo2020}. The emphasis on shared humanity inherent in Ubuntu will enable stakeholders to rethink how they develop and deploy educational AI, shifting the focus towards holistic systems that respect and support every person as a whole individual. Additionally, the collaborative nature of Ubuntu will transform the learning environment from an individualistic pursuit to a shared responsibility, a crucial shift for building community and trust in distance or online learning contexts \cite{Suliman2024}.
While the concept of relational trust offers a compelling normative alternative to prevailing models of AI trust, some argue that the ambiguity of relational theories, such as  Ubuntu, limits its practicality \cite{Mangaroo-Pillay2022}. We emphasise the importance of clear and actionable steps guided by experts are crucial for the operationalisation of relational concepts. Our trust by design principles foster genuine trust through transparency, long-term collaboration, and respect for all stakeholders. Finally, sustaining trust in AI requires inclusive deployment strategies and continuous participation of stakeholders, ensuring that trust is actively shaped between communities and those who develop AI systems \cite{Afroogh2024}. By operationalising these principles across the AI lifecycle, developers can foster trust, ensure ethical alignment, and build systems that truly serve the communities for which they are designed.

\section{Discussion}
In this paper, we propose principles to foster long-term trust in AI systems, grounded in relational ethics and informed by African philosophical traditions. Our approach challenges dominant paradigms that reduce trust to mere trustworthiness, defined narrowly in terms of legality, robustness, and technical reliability. While such criteria are essential, they are insufficient for cultivating genuine trust, especially in contexts where AI systems interact with human lives in morally significant ways. We posit that the normative conception of trust from relational ethics, emphasising bidirectional responsibilities, where the trustee must meet expectations and the trustor must give and accept vulnerabilities, is integral to genuine trust in AI.
In the philosophy of trust literature, the dominant conception of trust is unidirectional, focusing on higher levels of moral responsibility on the trustee \cite{ewuoso2023black}. This conception of trust is also predominantly applied to trust in AI or trustworthy AI principles. Compared to the relational normative conception of trust, we argue that treatments of trust in the existing literature are insufficient to create holistic trust. This is further compromised by the fragmentation of literature about trust in relation to AI. Technical literature focuses on transparency, accountability and privacy \cite{Hagendorff2020}. Yet, ethical principles and guidelines associated with trust can conflict  \cite{Thiebes2020}, For example, trade-offs between algorithmic performance and transparency, and performance versus privacy. Given such tensions, collaborative efforts are needed to address these tensions, thus improving trust in AI.
Philosophical debates increasingly critique the inadequacy of current trust frameworks that conflate trust with trustworthiness. As Durán and Pozzi\cite{Durán2025} argue, trust involves more than epistemic reliability; it requires an extra factor such as moral commitment, goodwill, and responsibility. This distinction is crucial in domains like healthcare, where trust in physicians is not solely based on their technical competence but also on their ethical commitment to beneficence and non-maleficence. Similarly, trusting AI systems in healthcare demands more than transparency or accuracy; it requires moral responsiveness and contextual accountability.
Trust is complex and requires a better understanding of the fundamental contributions of trust. To this end, Lukyanenko, Maass and Storey\cite{Lukyanenko2022} proposed a foundational trust framework inspired by Luhmann\cite{Luhmann2018} to understand the nature of trust in AI. Their framework has limitations, e.g., it assumes a non-reciprocal interaction between a trustee and a trustor. However, trust is bidirectional, grounded in social interactions with the obligation of reciprocity\cite{ewuoso2023black}. According to African relational ethics, reciprocity requires a willingness to be vulnerable and to accept vulnerability. This willingness fosters cooperation and enables trust. People are unlikely to cooperate or be vulnerable with those they do not trust \cite{Heyns2015}.
Relational ethics, particularly those rooted in African philosophical traditions, offer a compelling alternative. These traditions emphasise bidirectional trust, where both the trustor and trustee engage in mutual vulnerability and responsibility. As Ewuoso highlights, African relational ethics view trust as a social and moral obligation grounded in reciprocity, dignity, and communal well-being \cite{ewuoso2023black}. This perspective reframes trust not as a static attribute of systems but as a dynamic process of relationship-building. Our principles is thus centred on sustained collaboration with communities and guided by four relational principles:
Communitarianism: AI systems should be designed to benefit all members of society, especially marginalised groups, aligning with the African ethic of Ubuntu.
Respect for Persons: upholding dignity and moral worth in all interactions with AI.
Integrity: ensuring consistency, honesty, and ethical coherence in AI development and deployment.
Design Publicity: making AI systems explainable and transparent from the outset, including clear articulation of goals, algorithmic logic, and design motivations. 
While explainability is widely recognised as a trust-enhancing feature \cite{FerrarioLoi2022}, we argue that involving communities in shaping the justifications behind AI decisions deepens this trust. For instance, in healthcare, patients and clinicians should co-determine what constitutes a “reasonable explanation” for diagnostic or treatment recommendations. This participatory approach aligns with relational ethics and enhances moral legitimacy.
\subsection{Limitations}
The main limitations of our work is that we present a theoretical model and illustrate practical steps towards operationalisation without operational metrics for evaluating trust. Future research should develop tools that assess trust beyond accuracy and safety, incorporating relational and experiential dimensions. Also, the study does not encompass all perspectives on trust. Empirical research from diverse positionalities is needed to understand how AI systems challenge trust across different sectors and societies. For example, Gillespie et al.\cite{Gillespie2023} found that public trust in AI varies significantly across cultures and is deeply influenced by lived experiences. Similarly, a study in 157 Africans with educational interest in AI also found that trust varies across application areas \cite{amugongo2025trust}. Ultimately, building trust in AI requires more than technical robustness; it demands nuance in acting ethically, cultural sensitivity, and sustained engagement with communities. By incorporating African relational ethics and engaging with global majority perspectives, this paper contributes to a more inclusive and equitable discourse on AI trust. It invites the field to embrace alternative epistemologies that reflect the pluralism of human values and experiences, thereby enriching the global conversation on trustworthy AI.

\section{Conclusion}
In this paper, we propose principles that enable trust by design, inspired by relational ethics, to contribute to the discourse on trust in AI. Our approach not only provides a relational understanding of trust but also provides examples for developers or organisations developing AI systems that foster genuine trust. We introduced four principles: communitarianism, respect for others, integrity, and design publicity. These principles emphasise long-term collaboration, enabling the development of enduring relationships between communities and developers. By focusing on long-term collaborations, we advocate for a trust-by-design approach that ensures the meaningful involvement of those impacted by AI technologies in decision-making processes. In a nutshell, communitarian principles aim to balance individual and community interests, ensuring AI serves the best interests of everyone, especially marginalised groups. The principle of respect for others emphasises rights-respecting by design, where the inherent dignity of every person is upheld. Integrity promotes moral and ethical conduct, while design publicity encourages transparency by clearly defining system goals and their justifications.
Trust is a complex phenomenon that needs to be studied from different perspectives. While we do not claim that relational ethics provides all the answers to trust in AI, we posit that a relational view offers valuable principles for rethinking trust in AI, centring AI trust in long-term partnerships and collaboration with communities. Ensuring that AI systems respect relationships, autonomy, and the dignity of all persons. Additionally, emphasising transparency, inclusivity, and cultural sensitivity, we aim to foster genuine user trust in AI. 

\printbibliography

\end{document}